\documentclass[aps,prd,amsmath,amsymb,amsfonts,floatfix, twocolumn,eqsecnum,superscriptaddress,nofootinbib,10pt]{revtex4-2}

\usepackage{graphicx}
\usepackage[dvipsnames, usenames]{xcolor}
\definecolor{linkcolor}{rgb}{0.0,0.3,0.5}
\usepackage[colorlinks=true,allcolors=linkcolor]{hyperref}
\usepackage[normalem]{ulem} %for \sout

\newcommand{\PN}{\mathrm{PN}}
\newcommand{\PT}{\mathrm{PT}}
\newcommand{\Stop}{\mathrm{st}}

\newcommand{\UVA}{Department of Physics, University of Virginia, P.O.~Box 400714, Charlottesville, Virginia 22904-4714, USA}

\begin{document}

\title{Hybrid model for inspiral-merger-ringdown gravitational waveforms from comparable-mass, nonspinning binary black holes}

\author{Nur E.\ M.\ Rifat}
\email{nurrifat@virginia.edu}
\affiliation{\UVA}

\author{David A.\ Nichols} 
\email{david.nichols@virginia.edu}
\affiliation{\UVA}

\author{Kent Yagi} \email{kyagi@virginia.edu}
\affiliation{\UVA}

\begin{abstract}
Gravitational waves from comparable-mass binary-black-hole mergers are often described in terms of three stages: inspiral, merger and ringdown. 
Post-Newtonian and black-hole perturbation theories are used to model the inspiral and ringdown parts of the waveform, respectively, while the merger phase has been modeled most accurately using numerical relativity (NR).
Nevertheless, there have been several approaches used to model the merger phase using analytical methods. 
In this paper, we adapt a hybrid approximation method that applies post-Newtonian and black-hole perturbation theories at the same times in different spatial regions of a binary-black-hole waveform (and which are matched at a boundary region with prescribed dynamics).
Prior work with the hybrid method used leading post-Newtonian theory and the perturbation theory of nonrotating black holes, which led to errors during the late inspiral and disagreement with the dominant quasinormal-mode frequency extracted from NR simulations during the ringdown. 
To obtain a better match with NR waveforms of binary-black-hole mergers, we made several phenomenological modifications to the hybrid method.
Specifically, to better capture the inspiral dynamics, we use the effective-one-body method for modeling the trajectory of the boundary between the two spatial regions.
The waveform is determined by evolving a Regge-Wheeler-Zerilli-type equation for an effective black-hole perturbation theory problem with a modified Poschl-Teller potential. 
By tuning the potential to match the dominant quasinormal-mode frequency of the remnant black hole and also optimizing the boundary data on the matching region, we could match NR waveforms from nonspinning, comparable-mass binary black holes with mass ratios between one and eight, with a relative error of order $10^{-3}$.
\end{abstract}

\maketitle

\tableofcontents

\section{\label{sec:introduction} Introduction}

The LIGO, Virgo and KAGRA (LVK) detectors have been opening up a new field of gravitational-wave (GW) physics, which can probe general relativity (GR) for strongly gravitating systems, such as binary-black-hole (BBH) mergers. 
To date, more than 90 events have been detected during the LVK's first three observing runs, the majority of which are BBHs~\cite{LIGOScientific:2018mvr,LIGOScientific:2020ibl,LIGOScientific:2021djp}. 
During the longer fourth observing run, over 200 detection candidates have been announced~\cite{web:GraceDB}, and 128 compact binary coalescences from the first third of this observing run have now been confirmed~\cite{LIGOScientific:2025slb}.
Many of the searches or parameter estimation pipelines~\cite{Usman:2015kfa,Biwer:2018osg,Messick:2016aqy,Ashton:2018jfp,lalsuite} and the tests of general relativity~\cite{Berti:2018cxi,Berti:2018vdi,LIGOScientific:2019fpa,LIGOScientific:2020tif,LIGOScientific:2021sio,Yunes:2025xwp} rely upon waveform models of BBH mergers that faithfully represent the signal and are fast to evaluate.
This has led to a large number of analytical and numerical methods to compute gravitational waveforms and develop such waveform models. 

The most accurate waveforms come from numerical relativity (NR) simulations (see, e.g.,~\cite{Baumgarte:2010ndz}), because NR discretizes the full nonlinear Einstein equations in three spatial dimensions.
The most accurate of the NR waveforms are computed using Cauchy characteristic extraction (see, e.g.,~\cite{Winicour:2008vpn}).
The largest catalogs of BBH mergers contain hundreds~\cite{Jani:2016wkt} or thousands~\cite{Healy:2022wdn,Scheel:2025jct} of waveforms.
Running NR simulations (including those that use characteristic extraction methods) are computationally expensive, however. 
For example, the nearly four-thousand simulations in the catalog described in~\cite{Scheel:2025jct} took $4.8\times 10^8$ core-hours of computation time.
Models that are faster to evaluate than NR are essential for analyzing and interpreting GW observations.
In addition, interpreting the spacetime dynamics of NR simulations is nontrivial given their complexity, lack of symmetry, and dynamical spacetime coordinates.

\subsection{Discussion of waveform modeling methods}

One fruitful approach to constructing faster waveform models has been constructing NR waveform surrogates (see, e.g.,~\cite{Field:2013cfa,Field:2025isp}).
The NR surrogates are data-driven models, in the sense that they take existing NR simulations and construct interpolating functions on the intrinsic parameter space of BBH mergers (e.g., masses, spins and eccentricity).
The waveform from the surrogate model, when evaluated within the domain of validity of the model, reproduces the NR waveform with an accuracy comparable to the estimates of the numerical error in the NR simulations themselves.
The basis of interpolating functions used in the surrogate generally does not map to the functional expansions of waveforms used common analytical approximation methods, such as the post-Newtonian (PN) or multipolar post-Minkowskian (MPM) approximations~\cite{Blanchet:2013haa} or black-hole perturbation (BHP) theory (see, e.g.,~\cite{Berti:2009kk,Berti:2025hly}).
Thus, it is nontrivial to connect the insights from the surrogate model to those made from the analytical methods (e.g., PN and BHP theories).

Analytical methods can compute the waveform during portions of the signal, but they tend to be limited in their domain of applicability so that they cannot be used on their own to model all the stages of inspiral, merger and ringdown (IMR) during a BBH coalescence.
However, combining analytical results with phenomenological methods has been more successful in waveform modeling.
In some cases the analytical models also provide a simplified description of a BBH coalescence that helps interpret the BBH dynamics and the generation of gravitational waves.
An early example of this method is the close-limit approximation~\cite{Price:1994pm}, which used BHP theory much earlier into the merger (or even the late inspiral) than it would be expected to be valid.
This made it possible to obtain estimates of the GWs emitted during the merger phase using analytical data that approximates the infall of two black holes.
The success of this method has been understood in terms of the nonspherical perturbations in gravitational collapse~\cite{Price:1971fb}.
The subsequent work of the Lazarus project determined the limits of how early the close-limit approximation could be used in the merger phase~\cite{Baker:2001sf}.

A separate question that has been investigated is how late in the inspiral PN and MPM methods can be applied, and whether there are approaches that can push the applicability of PN methods later into the inspiral.
The effective-one-body (EOB) method~\cite{Buonanno:1998gg,Buonanno:2000ef} has done this by performing Pad\'e resummation methods (among other techniques) to extend PN results past the inspiral stage into the plunge stage.
This allows the binary dynamics to be evolved sufficiently close to the merger that a full IMR waveform can be constructed from the inspiral-plunge waveform (obtained directly from the EOB binary dynamics) by adding a superposition on quasinormal modes (QNMs) to the waveform.
A recent version of the EOB model, called SEOBNRv5PHM, takes information from NR and BHP (including second-order self-force calculations) to produce waveforms for precessing BBH systems that are sufficiently fast and accurate that they can be used for GW parameter estimation with the events observed by the LVK collaboration~\cite{Ramos-Buades:2023ehm}.

There are also phenomenological waveform models that take input from PN or BHP theory, but do not attempt to extend the validity of either method towards the merger.
Instead, they construct an interpolation between the two regions during the merger phase by directly fitting the NR waveform with elementary functions and enforcing a degree of continuity between the PN inspiral and QNM ringing in the waveform.
These IMRPhenom family of waveforms were originally developed in the frequency domain for the dominant quadrupole spherical-harmonic mode (e.g.,~\cite{Ajith:2007kx,Hannam:2013oca,Husa:2015iqa,Khan:2015jqa,Pratten:2020fqn}), though they have subsequently been extended to include subleading multipoles~\cite{Garcia-Quiros:2020qpx,Pratten:2020ceb} and to be constructed in the time domain~\cite{Estelles:2020osj,Estelles:2020twz}.
The IMRPhenomXPHM model~\cite{Pratten:2020ceb} is also used to analyze GW events from the LVK collaboration~\cite{lalsuite}, for example.

\subsection{Discussion about the ``hybrid method''}

The NR surrogate, SEOBNR and IMRPhenom waveforms are all successful at modeling the waveform from BBH mergers using techniques that are significantly more efficient computationally than running an NR simulation.
However, these techniques focus primarily on the waveform rather than on the spacetime dynamics that produce the GWs during the BBH coalescence.
An approximation method that attempted to both model the gravitational waveform and understand the underlying spacetime dynamics was introduced in~\cite{Nichols:2010qi,Nichols:2011ih}, and described as the ``hybrid method.''
The hybrid method used PN theory later and BHP earlier with respect to the merger phase by applying these approximation methods not in all of space at a given set of times, but simultaneously within different spatial regions (this differs from EOB or the close-limit approximation, which apply the approximation methods throughout all of space at a given time).
The spacetime dynamics resembled those in Price's model of stellar collapse~\cite{Price:1971fb}.
Rather than a collapsing star inside a perturbed black-hole spacetime, the hybrid method replaced the stellar interior with a PN description of an inspiraling BBH.
The inspiral waveform was sourced directly by the PN spacetime generating perturbations in the BHP spacetime exterior, whereas the merger and subsequent ringdown phase of the waveform in the hybrid model arose from the PN-generated perturbations interacting with, and subsequently getting hidden behind, the peak of the effective potential in the exterior BHP theory spacetime. 

In addition to providing this simple effective description of the spacetime during a BBH merger, the hybrid method was able to capture the qualitative features of the full IMR signal produced by numerical relativity, but it did not match the waveform as quantitatively as EOB does, for example.
Given that the results in~\cite{Nichols:2010qi,Nichols:2011ih} did not quantitatively reproduce the waveform, it was not apparent if this picture of PN perturbations interacting with an effective potential did indeed capture the full complexity of a BBH coalescence.
Specifically, it would be beneficial to determine if there are modifications to the hybrid method that can allow it to produce waveforms that quantitatively match the full IMR BBH waveforms while keeping the qualitative description of the spacetime dynamics similar.

In this paper, we investigate the types of modifications to the hybrid method that are required to fit the IMR waveforms from nonspinning BBH mergers with comparable, but unequal mass ratios.
To more precisely match the GW phase during the inspiral, we found it helpful to use EOB dynamics to prescribe the evolution of the timelike boundary from which the boundary data generates the waveform.
To better match the amplitude during the inspiral, we included tunable PN-motivated coefficients to rescale the boundary data.
There are other phenomenological modifications to the hybrid method that we needed to make to better match the merger and ringdown stages in the waveform.
The detailed description of these changes will be given in subsequent parts of the paper, but, in brief, they  include tuning the effective potential of the BHP theory used in the exterior (so as to better match the dominant ringdown frequency of the final black hole) and altering the late-time trajectory of the boundary region (so as to extend the EOB inspiral-plunge dynamics to later times).

\begin{figure*}
    \centering
    \includegraphics[width=\textwidth]{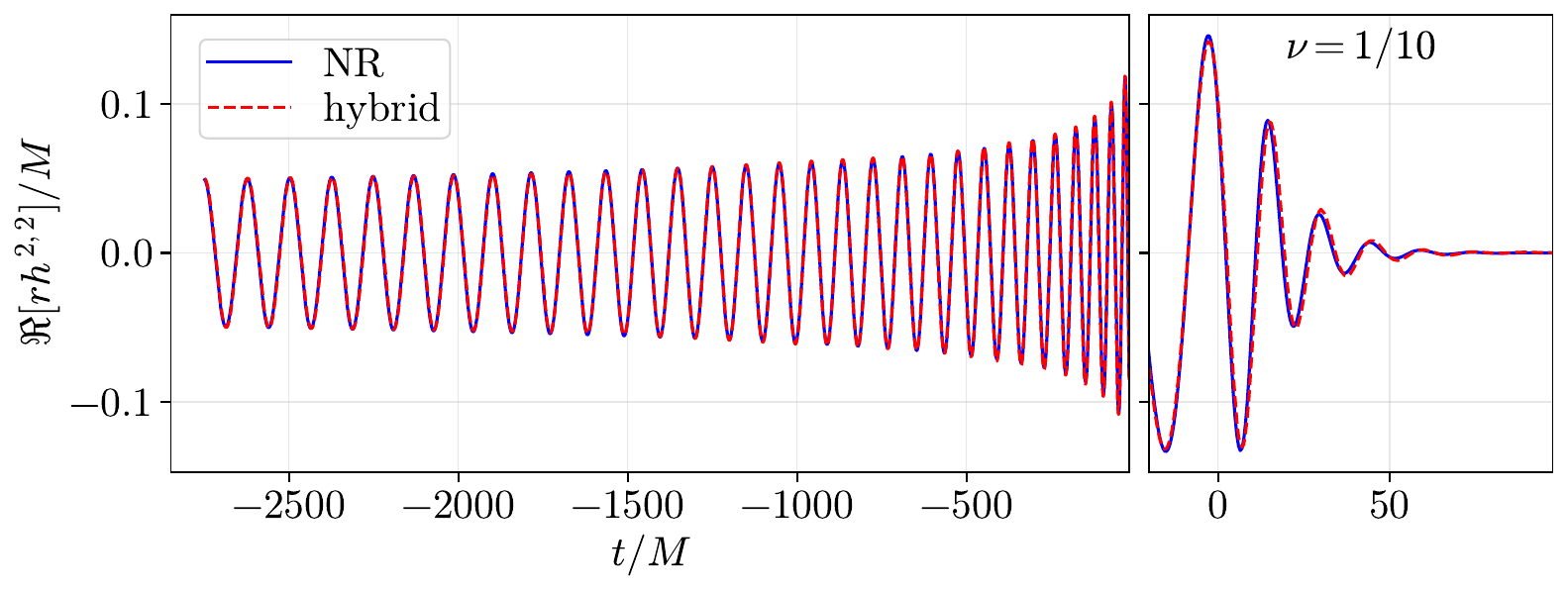}
    \caption{\textbf{Comparison of the hybrid-method and NR waveforms}:
    The red dashed curve shows the hybrid waveform, while the blue solid curves depict the NR waveform for the symmetric mass ratio $\nu = 1/10$ (where $\nu$ and $rh^{2,2}/M$ are defined in the text).
    The left panel shows the waveform during the inspiral stage, and the right one shows the waveform during the merger and ringdown stages.
    More details about how the hybrid waveform was computed will be given in the subsequent sections of this paper.}
    \label{fig:hyb_nr1}
\end{figure*}

With these modifications, we find that we can calibrate the hybrid method to match NR waveforms of nonspinning BBH systems with mass ratios from one to eight. 
As an example, we show a comparison between a waveform generated with the hybrid method and one from an NR surrogate model as the red-dashed and solid-blue curves, respectively, in Fig.~\ref{fig:hyb_nr1}. 
The left panel is a comparison of the inspiral stage and the right is a zoom in of the merger and ringdown stages of the waveform.
The waveform comes from nonspinning BBH with symmetric mass ratio $\nu = 1/10$, where $\nu = m_1 m_2/M^2$, $m_1$ is the primary's mass, $m_2$ is the secondary's mass, and $M=m_1+m_2$ is the total mass.
The waveform plotted is $r h^{2,2}/M$, where $r$ is the luminosity distance and the superscript ``2,2'' refers to just the $l=2$, $m=2$ spin-weighted spherical decomposition of the gravitational-wave strain $h$ (into spin weight $s=-2$ harmonics).
Units with $c=G=1$ are being used.

\subsection{Organization of this paper}

We briefly summarize the organization of the remainder of this paper.
In Sec.~\ref{sec:overview}, we review the essential elements of the hybrid method from the papers~\cite{Nichols:2010qi,Nichols:2011ih}.
Section~\ref{sec:modifications} is a discussion of the phenomenological modifications that we make to the hybrid method so that it generates gravitational waveforms that better match those from NR simulations.
We introduce four calibration parameters through these modifications, and how we calibrate the free parameters is discussed in Sec.~\ref{sec:calibration}.
Comparisons of the calibrated, hybrid-method waveforms with those from NR are given in Sec.~\ref{sec:GW}.
We conclude in Sec.~\ref{sec:conclusions} and present a few supplemental results in the Appendix. 
We use the geometric units of $c=G=1$ throughout the paper.

\section{\label{sec:overview} Review of the hybrid method} 

Here we summarize some aspects of the hybrid method~\cite{Nichols:2010qi,Nichols:2011ih}.
We provide just a short overview of the general methodology, as we will make several modifications to the approach of~\cite{Nichols:2010qi,Nichols:2011ih}, which will be described in Secs.~\ref{sec:modifications} and~\ref{sec:calibration}.
Further details about the original formulation of the hybrid method is given in~\cite{Nichols:2010qi,Nichols:2011ih}.

As described in Sec.~\ref{sec:introduction}, the hybrid method of~\cite{Nichols:2010qi,Nichols:2011ih} used PN and BHP theories at the same times, in limited regions of space, rather than in all of space up to or after a given time (the more typical usage).
Specifically, BHP theory was used on the exterior of a prescribed worldtube and PN theory was used in the interior of the worldtube.
The worldtube's size evolved with the orbital separation of the binary.
The spacetime metrics of the PN interior and BHP exterior were matched on this timelike worldtube (so that it was also referred to as a ``matching region'' or ``matching shell'').
The precise location of the matching region was chosen so that either both theories are valid at a certain level of accuracy or any inaccuracies in the matching procedure would not affect the GWs that were generated by the data on the matching region. 

The hybrid method ultimately reduces the computation of IMR waveforms from BBH mergers to solving a BHP theory problem in which no incoming waves were assumed to enter from past null infinity, and where the value of the Zerilli function on the timelike worldtube was prescribed by PN-motivated perturbations.
We summarize the quantitative details of this method in the next two subsections.

\subsection{\label{sec:matching_metric} Spacetime regions and matching}

Our notation follows that of~\cite{Nichols:2010qi,Nichols:2011ih}, where in the interior PN region, we will denote the coordinates with uppercase variables $(T, R, \Theta, \Phi)$ in a spherical-polar coordinate system.
We use lower-case letters $(t, r, \theta, \phi)$ for the Schwarzschild coordinate system in the exterior BHP spacetime.

Inside the worldtube, the leading Newtonian metric of a nonspinning BBH can be written in terms of the Newtonian potential, $U_N$.
For modeling the quadrupole gravitational waves, the Newtonian potential can be expanded to quadrupole order in the center-of-mass frame, so that it is given by
\begin{align} \label{eq:UN}
U_N \approx \frac{M}{R} + U_N^{(l=2)} ,
\end{align}
where $M$ is the total mass
The $l=2$ part of the potential can be written in terms of the following sum of spherical harmonics:
\begin{align} \label{eq:UNl2}
U_N^{(l=2)} = \sum_{m=-2}^{m=2} U_N^{2,m}(T,R)\,Y_{2,m}(\Theta, \Phi) .
\end{align}
In spherical polar coordinates constructed from the Cartesian harmonic coordinates $(X,Y,Z)$ using the flat-space relationships, the PN metric is given by 
\begin{align} \label{eq:ds2pn_sph}
\begin{split}
ds^2_{\PN} \approx & -\left(1-\frac{2M}{R} - 2U_N^{(l=2)}\right)dT^2 \\  
&+ \left(1+\frac{2M}{R} + 2U_N^{(l=2)}\right)(dR^2 +R^2d^2\Omega) \, .
\end{split}
\end{align}

The hybrid method assumed that the exterior metric was given by a perturbed Schwarzschild black hole with the metric
\begin{align} \label{eq:ds2sch}
    ds^2 \approx & -\left(1-\frac{2M}{r} \right)dt^2+\left(1-\frac{2M}{r}\right)^{-1}dr^2+r^2d^2\Omega \nonumber \\ 
    &+h_{\mu\nu}^{(l=2)} dx^{\mu}dx^{\nu} \, .
\end{align}
Here $h_{\mu\nu}^{(l=2)}$ is a quadrupolar metric perturbation. 
Given the metric in Eq.~\eqref{eq:ds2pn_sph}, it will be sufficient to assume that $h_{\mu\nu}^{(l=2)}$ is electric-parity [namely, it transforms as $(-1)^l$ under parity] so that just the Zerilli~\cite{Zerilli:1970wzz} perturbations are needed---the magnetic-parity Regge-Wheeler~\cite{Regge:1957td} perturbations will not be required.
Because we will match with a PN perturbation that is not in Zerilli gauge, however, we do not require that the Schwarzschild metric perturbation is in Zerilli gauge.
Nevertheless, we will use a notation similar to that of Regge and Wheeler~\cite{Regge:1957td} and Zerilli~\cite{Zerilli:1970wzz} for the metric perturbations: 
\begin{subequations}
\begin{align} \label{eq:sch_perts}
h_{tt}^{(l=2)} & = \sum_{m=-2}^{m=2} H_{tt}^{2,m}(t,r)\,Y^{2,m}(\theta,\phi) \, , \\
h_{tr}^{(l=2)} & = \sum_{m=-2}^{m=2} H_{tr}^{2,m}(t,r)\,Y^{2,m}(\theta,\phi) \, , \\
h_{rr}^{(l=2)} & = \sum_{m=-2}^{m=2} H_{rr}^{2,m}(t,r)\,Y^{2,m}(\theta,\phi) \, , \\
h_{\theta \theta}^{(l=2)} & = r^2 \sum_{m=-2}^{m=2} K^{2,m}(t,r)\,Y^{2,m}(\theta,\phi) \, , \\
h_{\phi \phi}^{(l=2)} & = h_{\theta \theta}^{(l=2)} \,\sin^2\theta \, .
\end{align}
\end{subequations}
The hybrid method of~\cite{Nichols:2010qi,Nichols:2011ih} matches the interior PN and exterior BHP metric at a fixed radius (the matching shell or worldtube). 
To do so, it must relate the harmonic coordinates used in the interior with the Schwarzschild coordinates used in the exterior.
First, for radii $r \gg M$, it uses the transformation from Schwarzschild to isotropic coordinates, which is of the form $r = R-M$ (to leading order in $M/r$), and with $t$, $\theta$, and $\phi$ unchanged.
After this transformation, the Schwarzschild metric takes the form
\begin{align} \label{ds2sch_approx}
    ds^2 \approx & -\left(1-\frac{2M}{R}\right)dt^2+\left(1-\frac{2M}{R}\right)(dR^2+R^2d^2\Omega) \nonumber \\ 
    &+h_{\mu\nu}dx^{\mu}dx^{\nu}\,,
\end{align}
where higher-order corrections in $M/R$ have been dropped.
By comparing the transformed Schwarzschild metric in Eq.~\eqref{ds2sch_approx} with the PN metric in Eq.~\eqref{eq:ds2pn_sph}, one can find that for the two metrics to match on a timelike worldtube of fixed $R$, the metric perturbation functions should be given by  
\begin{align} \label{eq:UN2Zerilli}
H_{tt}^{2,m} = H_{rr}^{2,m} = K^{2,m} = -2U_N^{2,m} , \qquad H^{2,m}_{tr} = 0 .
\end{align}
Because the quadrupole moment of the Newtonian potential of the binary along the matching region can be computed straightforwardly, it is used to determine the boundary data for the BHP problem.
This is reviewed in the next subsection.

\subsection{\label{sec:evolution_equation} Boundary data and evolution equations}

Because the metric-perturbation functions on the boundary, given in Eq.~\eqref{eq:UN2Zerilli}, are not in Zerilli gauge, it is convenient to compute the gauge-invariant Zerilli function from them.
After prescribing a location for the boundary (matching) worldtube, the Zerilli function can be evolved using the PN-inspired Zerilli boundary data on the matching worldtube and a no-incoming-wave condition from past null infinity.

Given the definition of the Zerilli function~\cite{Zerilli:1970wzz} (see also, e.g.,~\cite{Martel:2005ir,Ruiz:2007yx}) and the relations in Eq.~\eqref{eq:UN2Zerilli}, it was shown in~\cite{Nichols:2010qi,Nichols:2011ih} that on the boundary, the $l=2$, $m$ multipole moments of the Zerilli function are given in terms of the Newtonian potential by 
\begin{align} \label{eq:zerilli}
    \Psi^{2,m} = &\frac{2r}{3}\biggl \{U_N^{2,m} + \frac{r-2M}{2r+3M} \nonumber \\ 
    &\times \biggl[\biggl(1-\frac{2M}{r}\biggr)U_N^{2,m} - r\partial_rU_N^{2,m}\biggr]\biggr\} \, .
\end{align}
Treating the binary as two point masses of mass $m_1$ and $m_2$ (and reduced mass $\mu$), with semimajor axis $A(t)$, and orbital phase $\alpha(t)$, the quadrupole $U_N^{2,2}(t)$ is
\begin{equation} \label{eq:UN22}
U_N^{2,2} = \sqrt{ \frac{3\pi}{10}} \frac{\mu A(t)^2}{R^3} e^{-2i\alpha(t)} \, .
\end{equation}
The hybrid method focused on the $l=2$, $m=2$ mode, because that is the dominant waveform mode for nonspinning binaries.
We do the same, and we will henceforth drop the superscript ``$2,2$'' on the Zerilli function.

Given the explicit expression for the $(2,2)$ moment of the Newtonian potential, this quadrupole can now be substituted into Eq.~\eqref{eq:zerilli} to obtain the expression for the gauge-invariant Zerilli function.
Performing the transformation $R = r + M$ and keeping the leading-order terms in $R$, gives the expression in~\cite{Nichols:2011ih}:
\begin{equation} \label{eq:zerilli_lead}
\Psi = 2\sqrt{ \frac{3\pi}{10}} \frac{\mu A(t)^2}{R^2} e^{-2i\alpha(t)}\,.
\end{equation}
To evaluate the expression in Eq.~\eqref{eq:zerilli_lead} on the boundary between the PN interior and BHP exterior, the radius of the shell must be specified.
In~\cite{Nichols:2010qi,Nichols:2011ih}, only equal-mass binaries were considered, and the location of the boundary shell was chosen to coincide with the orbital location of the two bodies in the PN spacetime, $A(t)/2$.
A natural generalization of this criteria for unequal-mass binaries is to choose it to be the location of the secondary (the further out of the two bodies in the center-of-mass frame), $(m_1/M)A(t)$, where $m_1 > m_2$ (and $M=m_1+m_2$ is the total mass).
Evaluating the Zerilli function in Eq.~\eqref{eq:zerilli_lead} on the shell $R = (m_1/M)A(t)$ gives
\begin{equation} \label{eq:zerilli_shell}
\Psi|_\mathrm{shell} = 2 \sqrt{ \frac{3\pi}{10}} \frac{m_2}{m_1} M e^{-2i\alpha(t)} .
\end{equation}

The Zerilli function on the timelike boundary worldtube and a no-incoming-wave condition on the initial value of the advanced time coordinate were argued in~\cite{Nichols:2010qi,Nichols:2011ih} to be consistent boundary and characteristic-initial-value data to specify an evolution problem for the Zerilli equation.
Following~\cite{Nichols:2010qi,Nichols:2011ih}, we use the light-cone coordinates $(u,v)$, which are defined by
\begin{equation} \label{eq:LCcoords}
u = t-r_* \, , \hspace{0.5cm} v=t+r_* , 
\end{equation}
where $r_*$ is the tortoise coordinate:
\begin{equation} \label{eq:r_star}
r_* = r +2M \log\left(\frac{r}{2M}-1 \right).
\end{equation}
In these null coordinates, the Zerilli equation satisfies the evolution equation
\begin{equation} \label{eq:RWZ}
\frac{\partial^2\Psi^{l,m}}{\partial u \partial v} + \frac{1}{4} V_l(r) \Psi^{l,m} = 0 .
\end{equation}
The Zerilli potential, $V_l(r)$ is given by
\begin{subequations}
\begin{equation} \label{eq:vl_zerilli}
V_{l}(r) = \left( 1-\frac{2M}{r} \right) \left[ \frac{\lambda}{r^2}-\frac{6M}{r^3}U_{l}(r) \right] ,
\end{equation}
where $ \lambda = l(l+1) $.
The function $U_l(r)$ is defined by
\begin{equation} \label{eq:ul_zerilli}
U_{l}(r) = \frac{ \Lambda (\Lambda + 2 ) r^2 + 3M (r-M) }{ (\Lambda r + 3m)^2 },
\end{equation}
where we have also defined
\begin{equation} 
\Lambda = (l-1)(l+2)/2 = \lambda/2 - 1.
\end{equation}
\end{subequations}
We use a second-order-accurate numerical method to evolve the Zerilli function (see, e.g.,~\cite{Gundlach:1993tp}).
We treat the region adjacent to the boundary using the method described in~\cite{Nichols:2011ih}.

Evolution equations for the separation $A(t)$ need to be provided to describe the evolution of the boundary region.
The trajectory $A(t)$ [an the orbital phase $\alpha(t)$] was given in~\cite{Nichols:2011ih} by a set of Hamiltonian equations, supplemented by a dissipative force, of the form
\begin{align} \label{eq:HamiltonsEqs}
    \dot A(t) = \frac{\partial H}{\partial p_A} , &  \qquad \dot \alpha(t) = \frac{\partial H}{\partial p_\alpha} ,\nonumber \\
    \dot p_A(t) = -\frac{\partial H}{\partial A} , &  \qquad \dot p_\alpha(t) = \mathcal F_\alpha  .
\end{align}
In~\cite{Nichols:2011ih}, the Hamiltonian was chosen to be the Hamiltonian for point-particle motion in the Schwarzschild spacetime, and the dissipative torque $\mathcal F_\alpha$ was obtained from a radiation-reaction potential that was computed self-consistently from the Zerilli function. 
However, because we will use the EOB Hamiltonian and radiation-reaction forces in the subsequent calculations, we do not review the details of the Hamiltonian and force used in~\cite{Nichols:2011ih}.

\section{\label{sec:modifications} Phenomenological modifications to the hybrid method} 

In this section, we summarize the main limitations of the hybrid method of~\cite{Nichols:2010qi,Nichols:2011ih} for producing waveforms that match the results from NR simulations, and we discuss the modifications to the hybrid method that we implement to improve the agreement between these waveforms. 
This section will cover the modifications to the trajectory of the matching shell, the boundary data on this shell, and potential of the effective BHP spacetime in its three respective subsections.
The details of the procedure to calibrate the free parameters that we add to hybrid method will be given in Sec.~\ref{sec:calibration}.

\subsection{\label{subsec:trajectory} Trajectory of the matching shell}

As noted in Sec.~\ref{sec:evolution_equation}, the hybrid model of~\cite{Nichols:2010qi,Nichols:2011ih} computed the trajectory of the matching shell by solving Eq.~\eqref{eq:HamiltonsEqs} where $H$ was the Hamiltonian for geodesic motion in the Schwarzschild spacetime. 
For inspirals, the radiation-reaction torque was computed via a Burke-Thorne potential~\cite{Burke:1970wx,Thorne:1969rba} that is consistent with the outgoing GWs in the BHP spacetime (and for head-on collisions, the radiation-reaction effects on the equations of motion were neglected).
For a nonspinning equal-mass BBH, the GW phase from the hybrid method agreed well with an NR waveform (for thousands of $M$), until the last few cycles of the inspiral.
The amplitude of the waveform agreed less well, however.
During the merger stage and ringdown stages, there were significant disagreements in both amplitude and phase.

Because the EOB framework is constructed to extend the validity of the PN expansion to the late-inspiral and plunge stages of a BBH coalescence, it is natural to consider if replacing the Schwarzschild Hamiltonian with the EOB Hamiltonian and the EOB radiation-reaction generalized forces could improve the agreement with numerical relativity.
We will adopt this approach in this paper.
Because we focus on nonspinning BBH systems in this paper, we use the early EOB model, EOBNRv2, which is described in~\cite{Pan:2011gk}.
This model is implemented in the LVK Algorithm Library (LAL) software suite~\cite{lalsuite} as part of the \textsc{LALSimulation} code base.
We extract the EOB trajectory from this implementation of the EOB model

There are several subtleties that will arise from using the EOB Hamiltonian to model the dynamics of the matching shell.
When the EOB method is used for waveform modeling, the EOB dynamics are truncated near the EOB light ring (to complete the waveform from the EOB trajectory, a superposition of QNMs is fit to the waveform computed from the EOB inspiral and plunge).
The hybrid method, however, requires a trajectory for the matching shell that continues for the duration of the simulation (i.e., including retarded times that correspond to the ringdown) rather than truncating at a fixed radius (and time).
In~\cite{Nichols:2010qi,Nichols:2011ih}, the Hamiltonian for point-particle motion in the Schwarzschild spacetime was used because, in Schwarzschild coordinates, plunging particles asymptote to the horizon exponentially in time, with a time constant determined by the black hole's surface gravity (see, e.g.,~\cite{Nichols:2010qi}).
In the Schwarzschild tortoise coordinate $r_*$, this implies that trajectory approached minus infinity linearly in time during the plunge.

The trajectory generated by the EOBNRv2 equations of motion allows the particle to approach zero radius rather than the horizon of the effective EOB spacetime (which for EOBNRv2 occurs at a negative radius).  
Thus, there was not an obvious way to use the EOB trajectory at all times and obtain dynamics of the matching shell that were qualitatively similar to those used in~\cite{Nichols:2010qi,Nichols:2011ih}.
To mimic the late-time approach to the horizon in~\cite{Nichols:2010qi,Nichols:2011ih} while still using the EOB dynamics at earlier times, we will stop the EOB dynamics at a time $t_\Stop$, which corresponds to a separation $A_\Stop \equiv A(t_\Stop)$ near the EOB light ring.
We will smoothly match a trajectory that rapidly approaches the horizon at times $t > t_\Stop$ and at separations less than $A_\Stop$.
We give the details of this fitting later in this section, after we make a few additional comments about the EOB radius and our choice of tortoise coordinate.

Although the evolution equations are written in terms of the separation $A$, we need to relate the EOB trajectory in the EOB coordinates to that of the exterior BHP spacetime, which was written in terms of the Schwarzschild coordinate $r$.
Using the EOB coordinate is advantageous, because as was shown in~\cite{Buonanno:1998gg}, the EOB effective spacetime metric agrees with the Schwarzschild metric up to 2PN corrections.
Because we are working at Newtonian order, then we do not need to perform the coordinate transformation to relate the coordinates of the PN interior to the BHP exterior, as in~\cite{Nichols:2010qi,Nichols:2011ih}.
Thus, we can use $A(t)$ or $r(t)$ interchangeably in the discussion below [e.g., we can refer to the stopping separation as $A_\Stop$ or $r_\Stop \equiv r(t_\Stop)$].

The evolution of the Zerilli function, however, is performed in the light-cone coordinates $t\pm r_*$.
Thus, it also is necessary to transform the location of the matching shell to these coordinates, which involves a choice of the tortoise coordinate.
Given the behavior of the EOB separation $A(t)$, discussed above, the trajectory of the shell in the EOB tortoise coordinate will also not have the qualitative behavior of the trajectory in~\cite{Nichols:2010qi,Nichols:2011ih}.
While we could again use the Schwarzschild tortoise coordinate, because we will be modifying the potential of the effective BHP problem in the exterior region (as we discuss in more detail in Sec.~\ref{subsec:modPotential}), we opt instead to use the tortoise coordinate $r_*$ of the remnant Kerr black hole formed after the merger.

Specifically, the Kerr tortoise coordinate can be expressed as the solution of the differential equation
\begin{equation}
    \frac{dr_*}{dr} = \frac{r^2 + a_f^2}{r^2 - 2M_f r + a_f^2} .
\end{equation}
The final mass $M_f$ and final spin parameter $a_f$ resulting from a BBH merger with symmetric mass ratio $\nu$ are obtained from the expressions used in the EOBNRv2 model~\cite{Pan:2011gk}, which we reproduce below:
\begin{subequations} \label{eq:remnantBHfits}
    \begin{align}
        \frac{M_f}{M} = {} &  1 + \left(\sqrt{\frac 89} - 1 \right) \nu - 0.4333\nu^2 - 0.4392\nu^3 , \\
        \frac{a_f}{M_f} = {} & \sqrt 12 \nu -3.87\nu^2 + 4.028\nu^3 .
    \end{align}
\end{subequations}
On the boundary, we will also write the tortoise coordinate as just $r_*$, given the equivalence of the EOB $A$ and $r$ at 1PN order, discussed above.

With this choice of tortoise coordinate, we can now discuss how we match the EOB trajectory to a trajectory that approaches the boundary at the desired rate.
We will denote the EOB trajectory in terms of the tortoise coordinate by $r_*^\mathrm{EOB}(t)$ and the late-time (LT) trajectory by $r_*^\mathrm{LT}(t)$. 
First, we stop evolving the EOB equations of motion at time $t_\Stop$ which is determined by the condition that $r^{\mathrm{EOB}} (t_\Stop) \equiv r^\mathrm{EOB}_\Stop = r_{+} + 0.04M$.
We use the notation $r_{+}$ for the radius of the outer horizon.
The radii of outer and inner horizons ($r_\pm$, respectively) of a Kerr black hole with mass $M_f$ and spin parameter $a_f$ [which are determined from the fitting functions in Eq.~\eqref{eq:remnantBHfits}] are
\begin{align} \label{rHp}
r_{\pm} = M_{f} \pm \sqrt{M_{f}^2 -a_{f}^2}\, .
\end{align}

In Ref.~\cite{Nichols:2010qi}, it was shown that in the tortoise coordinate, the plunging trajectory of a head-on merger approached a line of slope negative one as the matching shell approached the horizon.
The precise rate of the approach to the horizon for an inspiral was not determined in~\cite{Nichols:2011ih}.
In this paper, we make the ansatz that the late-time trajectory still approaches the horizon at a constant rate, which we empirically fix to be minus one half. 
While this choice is somewhat arbitrary, we found that we could smoothly match $r_*^\mathrm{LT}(t)$ to the EOB trajectory $r_*^\mathrm{EOB}(t)$ and, furthermore, this slope did not lead to artifacts in the waveform generated through our modified hybrid method.
Given our choice of $r^\mathrm{EOB}_\Stop = r_{+} + 0.04M$, we find that for all mass ratios that we consider, there is a time $t < t_\Stop$ where $\dot r_*^\mathrm{EOB} (t)$ equals $-1/2$, so that the first derivatives of the EOB and LT trajectories can be matched.
We define the time at which this occurs to be $t_\mathrm{half}$, and we write the late-time trajectory as
\begin{align} \label{rLT}
r_{*}^{\mathrm{LT}}(t) = -\frac 12 (t - t_\mathrm{half}) + r_{*,\mathrm{half}} \,.
\end{align}
The constant $r_{*,\mathrm{half}}$ is chosen so that the late-time trajectory is continuous with $r_*^\mathrm{EOB}$. 

Enforcing continuity at $t_\mathrm{half}$ would, by construction, make the trajectory continuous with a continuous first derivative.
To enforce a higher degree of continuity, we ``blend'' the EOB and LT trajectories using a hyperbolic-tangent tapering function.
Empirically, again, we find that a window of the form 
\begin{equation} \label{eq:Window}
H(t) =  \frac{1}{2} \left[ 1 + \tanh \left( \frac 4 M (t - t_\mathrm{half} + M) \right) \right]
\end{equation}
will allow us to smoothly match the two trajectories.
The window is applied symmetrically about the location $t_\mathrm{half} - M$ [the zero of the hyperbolic tangent function in Eq.~\eqref{eq:Window}].
The full boundary trajectory is given by
\begin{equation}
    r^{\mathrm{Full}}_{*}(t) = H(t) r_{*}^{\mathrm{LT}}(t) + \left[1 - H(t)\right] r^{\mathrm{EOB}}_{*}(t) .
\end{equation}
Note that within $\mp 2M$ of $t_\mathrm{half}-M$ the window function in Eq.~\eqref{eq:Window} is of order $e^{-16} \approx 10^{-7}$ (or one minus that value, respectively).
Thus, to good approximation, the EOB (respectively, LT) trajectory can be used for times less than (greater) than $\mp 2M$ of $t_\mathrm{half}-M$.

\begin{figure}[htb]
    \centering
    \includegraphics[width=\columnwidth]   
    {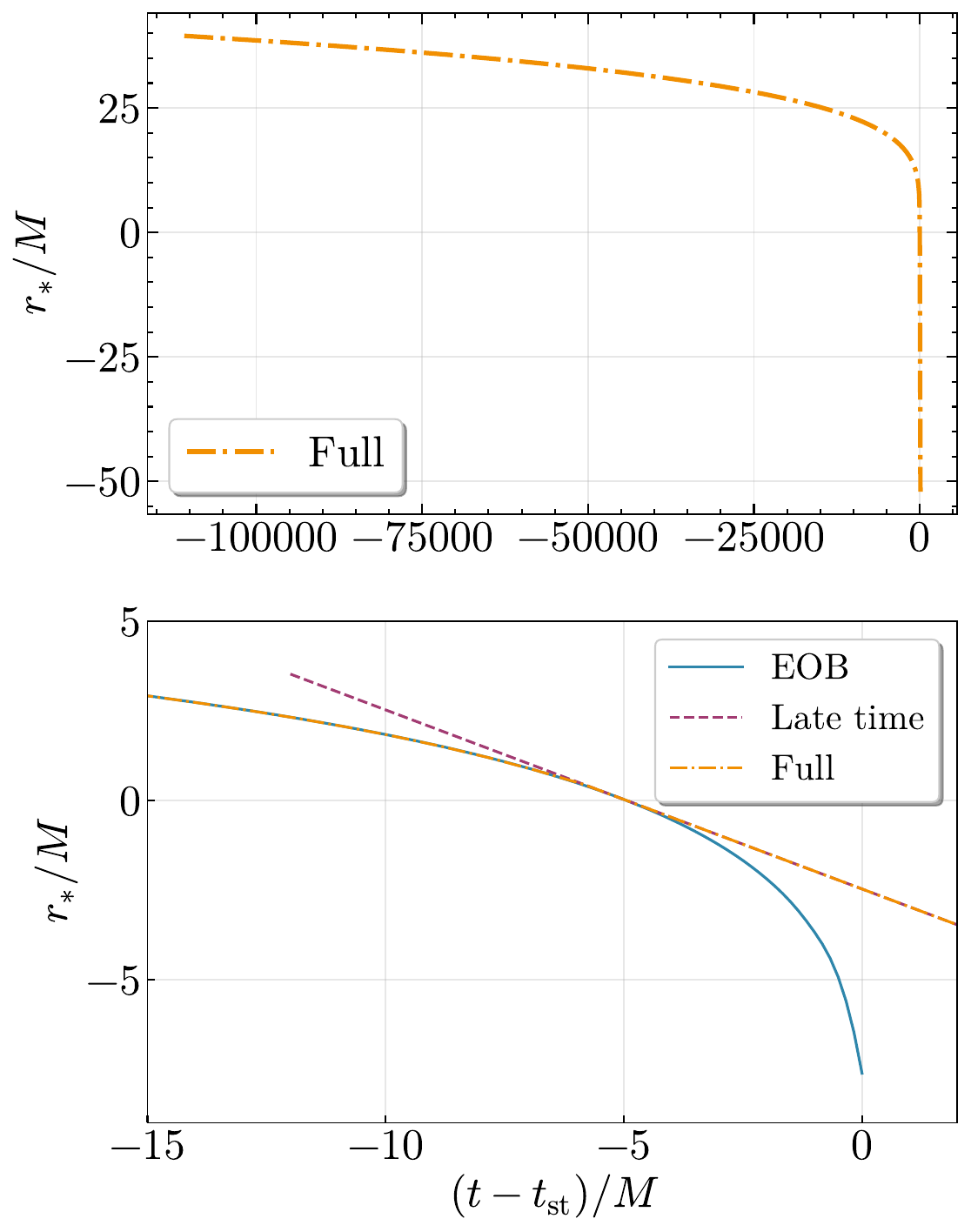}
    \caption{\textbf{Trajectory of the matching shell in the Kerr $r_*$ coordinate for an equal mass-ratio BBH}.
    In both panels, the trajectory corresponds to an equal-mass BBH system. 
    In the top panel, the orange dash-dotted curve shows full trajectory after EOB and LT trajectories are smoothly blended. 
    In the bottom panel, the solid blue curve shows the trajectory obtained from solving the EOB equations of motion. 
    It stops at a time $t_\Stop$, which we have defined to be $t_\Stop = 0$. 
    The maroon dashed line is the late-time trajectory, which is constructed to be the tangent to the EOB trajectory when it reaches a slope of $-1/2$.}
    \label{fig:trajectory}
\end{figure}

We show the boundary trajectory in Fig.~\ref{fig:trajectory}.
The top panel is the full trajectory shown over a timescale of order $10^5M$.
The bottom panel shows both EOB and LT trajectories over timescales of order $10 M$, and how they are smoothly blended around $t_\mathrm{half}$, where the two trajectories have equal slopes.

Finally, the EOB Hamiltonian also produces an evolution equation for the orbital phase $\alpha$.
When the EOB trajectory is truncated at the time $t_\Stop$, this will also stop the evolution of the orbital phase abruptly at a value which we denote $\alpha_\Stop \equiv \alpha(t_\Stop)$.
This requires that we have a method of evolving the orbital phase at times $t > t_\Stop$.
In the original hybrid method of~\cite{Nichols:2010qi,Nichols:2011ih}, the orbital phase smoothly approached a constant as the matching shell approached the horizon on a timescale determined by the inverse of the surface gravity of the black hole.
In our modification of the hybrid method, we will enforce this late-time behavior by fitting an exponential approach to a constant phase.
However, we will use the surface gravity of the remnant Kerr black hole rather than that of a Schwarzschild black hole.

Specifically, to enforce continuity of the $\alpha(t)$ and its first two derivatives at $t_\Stop$, we make an ansatz for $\alpha(t)$ when $t > t_\Stop$ of the form
\begin{equation} \label{eq:alpha_exp}
    \alpha(t) = \alpha_\Stop + \alpha_1 \left[1 - e^{-\kappa_H (t-t_\Stop)}\right] + \alpha_2 \left[1 - e^{-2\kappa_H (t-t_\Stop)}\right] .
\end{equation}
In Eq.~\eqref{eq:alpha_exp}, we have used the notation $\kappa_H$ for the surface gravity given by
\begin{align} \label{kappa}
\kappa_{H}= \frac{r_{+}- r_{-}}{r_{+}^2 + a_{f}^2} \, .
\end{align}
Enforcing continuity of the first and second derivatives of $\alpha(t)$ with the EOB orbital phase at $t_\Stop$ gives rise to linear equations that can be solved for the undetermined coefficients $\alpha_1$ and $\alpha_2$.
The form of the ansatz enforces continuity of the phase at $t_\Stop$ by construction.

\subsection{\label{subsec:boundary} Boundary data on the matching shell}

While using the EOB dynamics will make significant improvements in matching the GW phase of the hybrid method with that of NR during the inspiral, it will not address the discrepancy in the amplitude.
We find that this can be resolved by introducing two phenomenological parameters, which we denote by $C_{0\PN}$ and $C_{1\PN}$.

The coefficient $C_{0\PN}$ is used as an overall scaling to the Zerilli function on the boundary, which allows us to tune the amplitude of the hybrid-method waveform to better agree with that of NR during the earlier stages of the inspiral.
However, we found that allowing $C_{0\PN}$ to be a function of mass ratio $q = m_2/m_1$ is not sufficient to make the NR and hybrid waveforms agree during the late inspiral and merger.
If we add an effective 1PN correction to the boundary data on the matching shell of the form $M/R$ times a new coefficient $C_{1\PN}(q)$, then we are able to make the NR and hybrid waveforms agree to good accuracy.
Thus, we will use the following expression for the Zerilli function on the boundary:
\begin{align} \label{zerilli_shell3}
\Psi|_\mathrm{shell} = 2 q M e^{-2i\alpha(t)}
\left[ C_{0\PN}(q) - C_{1\PN}(q) \left(\frac{M}{R} \right) \right] .
\end{align}
We have absorbed the order-one coefficient $\sqrt{3\pi/10}$ in Eq.~\eqref{eq:zerilli_lead} into the coefficients $C_{0\PN}$ and $C_{1\PN}$, for convenience. 
The orbital phase $\alpha(t)$ and the orbital separation $A(t)$ are obtained from the trajectory described in Sec.~\ref{subsec:trajectory}. %\ky{Sec.~\ref{sec:evolution_equation}?}. 
We describe how we calibrate the coefficients $C_{0\PN}(q)$ and $C_{1\PN}(q)$ in Sec.~\ref{sec:calibration}.

\subsection{\label{subsec:potential} Modifications to the black-hole-perturbation potential} \label{subsec:modPotential}

In the head-on BBH collisions considered in~\cite{Nichols:2010qi}, it was reasonable to consider perturbations of a Schwarzschild black hole during the ringdown stage, because the BBH merger resulted in a nonspinning remnant.
This assumption was less well justified for the quasicircular inspirals of nonspinning BBHs considered in~\cite{Nichols:2011ih}, for which the remnant BH has a dimensionless spin of roughly $0.67$ (see, e.g.,~\cite{Scheel:2025jct}).
Thus, there was a significant discrepancy, during the ringdown, between the waveform from the hybrid method for inspirals in~\cite{Nichols:2011ih} and from NR simulations, because~\cite{Nichols:2011ih} continued to use a nonrotating BHP theory spacetime in the exterior.

A possible way to obtain the QNM frequency observed in the ringdown stage of NR simulations would be to use as the exterior BHP spacetime a perturbed Kerr black hole with a spin parameter consistent with the remnant mass predicted by NR.
Performing a Kerr BHP problem would be much more demanding technically and require significant modifications to the hybrid method.
Specifically, they would include the facts that (i) the Teukolsky equation, when written in the time domain after separating in the angular coordinate $\phi$, is a partial differential equation in two spatial dimensions rather than one; (ii) there is not an equivalent characteristic evolution algorithm as that in the Schwarzschild spacetime (e.g.,~\cite{Gundlach:1993tp}); (iii) the boundary would need to be an axisymmetric surface rather than a spherical shell, which would introduce considerable additional freedom in defining the matching region.

Instead, we take a purely phenomenological approach, in which we still consider a problem that takes a RWZ form, but the potential is modified from the RWZ potential in Eq.~\eqref{eq:vl_zerilli} so that the least-damped QNM of this potential matches that in the corresponding NR simulation.
In a sense, this is the most dramatic phenomenological modification to the hybrid method, as by modifying the potential, the effective Zerilli equation is no longer consistent with a gauge-invariant construction that satisfies the linearized Einstein equations on a black-hole background.
Nevertheless, we will continue to refer to the exterior spacetime as a modified BHP theory spacetime. 
The intent of this approach is, as discussed in Sec.~\ref{sec:introduction}, to understand the extent to which PN-inspired data interacting with an effective potential can represent the NR waveform during the late inspiral, merger and ringdown; we are not attempting to represent all aspects of the spacetime with the highest degree of fidelity.

The potential that we use is what we will refer to as a modified Poschl-Teller (mPT) potential. 
Poschl-Teller (PT) potentials have the property that they have a simple analytical form and their QNMs are known analytically.
However, we found it challenging to match the waveform from NR simulations using just this potential (despite the fact that it has been successful in modeling the Weyl scalar $\Psi_4$ in the backward-one-body method~\cite{McWilliams:2018ztb}).
Given that the PT potential is short range, and that the Zerilli potential falls off more slowly as $1/r^2$ (which is just radial scaling of the centrifugal potential in spherical coordinates in flat spacetime), we construct a potential that has a slower fall off. 
Our approach is to use the PT potential at radii smaller than the light ring of co-rotating equatorial orbits and to have the potential transition to one that is a polynomial in inverse powers of $1/r$ at radii larger than the light ring.
The location of the light ring $r_\mathrm{lr}$ for a Kerr black hole of mass $M_f$ and spin parameter $a_f$ is the intermediate root of the cubic polynomial equation
\begin{equation}
    r_\mathrm{lr} (r_\mathrm{lr}-3M_f)^2 - 4 M_f a_f^2  = 0 .
\end{equation}

\begin{figure}[htb]
    \centering
    \includegraphics[width=\columnwidth]{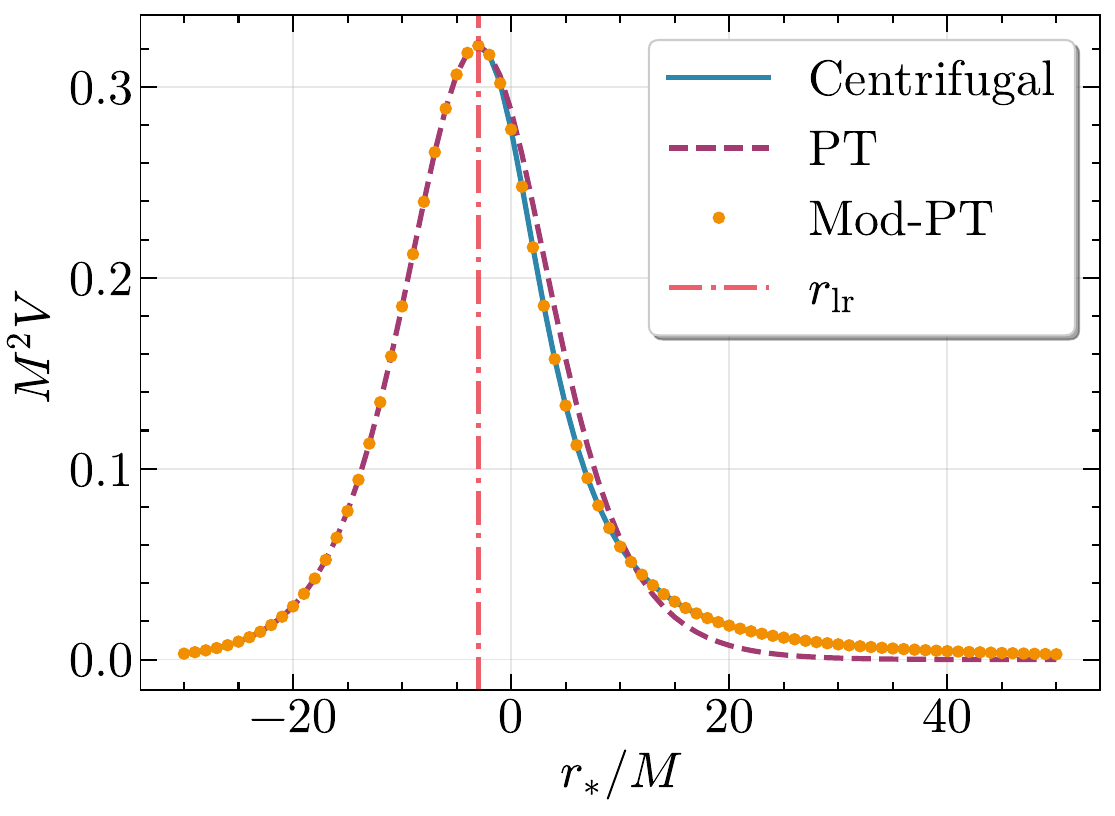}
    \caption{\textbf{The modified Poschl-Teller potential for an equal mass system}. 
    The full mPT potential is shown as a dotted orange curve. 
    It is constructed from a PT potential, the dashed purple curve, which is shown at all values of $r_*$. 
    The potential that is a polynomial in $1/r$ at $r \geq r_\mathrm{lr}$, which we label as the ``Centrifugal'' potential, is shown as a solid blue curve.
    The red vertical line shows where the matching between the two potentials takes place at $r_\mathrm{lr}$. 
    }
    \label{fig:mod_pt}
\end{figure}

Specifically, we write the mPT potential $V_\mathrm{mPT}(r)$ as
\begin{equation} \label{mod_pt}
    V_\mathrm{mPT}(r) = 
    \begin{cases} 
      V_\mathrm{PT} \, \mathrm{sech}^2\boldsymbol(\kappa_\mathrm{PT} [ r_{*}(r) - r_{*}(r_\mathrm{lr})] \boldsymbol) & r < r_\mathrm{lr} , \\
      \dfrac{l(l+1)}{r^2}+\displaystyle\sum_{k=3}^{5}\frac{C_k}{r^k} & r\geq r_\mathrm{lr} ,
    \end{cases}
\end{equation}
where the coefficients $C_k$ are obtained from solving the three linear equations that arise when enforcing continuity of the potential and its first two derivatives at $r=r_\mathrm{lr}$.
We leave the amplitude $V_\PT(q)$ and the inverse length scale $\kappa_\PT(q)$ of the PT potential as coefficients which we calibrate against NR simulations of different mass ratios.
The details of how we calibrate the coefficients are given in Sec.~\ref{sec:calibration}.
For illustrative purposes, we show the mPT potential, and its component parts, as a function of $r_*$ in Fig.~\ref{fig:mod_pt}.

\section{\label{sec:calibration} Calibrating the modifications to the hybrid method} 
    
For calibrating the hybrid waveforms we use six symmetric mass ratios ($\nu = 1/4$, $21/100$, $1/6$, $1/8$, $1/10$), where $\nu$ can be written in terms of the mass ratio $q=m_2/m_1$ as
\begin{equation}
    \nu = \frac{q}{(1+q)^2} .
\end{equation}
We use the NR surrogate model described in~\cite{Blackman:2015pia} as our NR waveform model against which we calibrate the hybrid model. 

For each of these values of the symmetric mass ratio, we run several simulations of the hybrid method by solving the modified RWZ equation
\begin{equation} \label{eq:modRWZ}
\frac{\partial^2\Psi}{\partial u \partial v} + \frac 14 V_\mathrm{mPT} \Psi = 0 
\end{equation}
for specific choices of the free parameters $C_{0\PN}$, $C_{1\PN}$, $V_\PT$ and $\kappa_\PT$.
While we use the modified PT potential $V_\mathrm{mPT}$ given in Eq.~\eqref{mod_pt} and the new boundary data in Eq.~\eqref{zerilli_shell3} with the modified shell location, the approach to solving Eq.~\eqref{eq:modRWZ} is the same as in~\cite{Nichols:2010qi}; further details can be found there and in~\cite{Nichols:2011ih}.

To optimize the parameters $C_{0\PN}$, $C_{1\PN}$, $V_\PT$ and $\kappa_\PT$, we will them to minimize an error function used in~\cite{Blackman:2015pia}.
The error is defined as the ratio of the $L^2$ norm of the difference of the hybrid and NR waveforms (the latter evaluated using the surrogate~\cite{Blackman:2015pia}) and the $L^2$ norm of the NR waveform:
\begin{equation}
\begin{aligned} \label{eq:cal_e}
\mathcal E[h^{2,2}_\mathrm{hyb},h^{2,2}_\mathrm{NR}] = \frac{\displaystyle\int_{t_0}^{t_f} dt \left|h_\mathrm{hyb}^{2,2}(t ; q)-h_\mathrm{NR}^{2,2}(t ; q)\right|^{2}}{2\displaystyle\int_{t_0}^{t_f}dt \left|h_\mathrm{NR}^{2,2}(t ; q)\right|^{2}}\,.
\end{aligned}
\end{equation}
The factor of two in the denominator is introduced to agree with the conventions for this error measure used in~\cite{Blackman:2017dfb}.

Because a simulation with the hybrid method takes on the order of a few minutes to complete, using a minimization algorithm to find the optimal values of the parameters $C_{0\PN}$, $C_{1\PN}$, $V_\PT$ and $\kappa_\PT$ that minimize Eq.~\eqref{eq:cal_e} would be computationally intensive.
To speed up the parameter tuning, we perform a sequential optimization of the potential parameters $V_\PT$ and $\kappa_\PT$ at fixed $C_{0\PN}$ and $C_{1\PN}$ to match the least-damped ringdown frequency of the remnant Kerr black hole.
With these optimized $V_\PT$ and $\kappa_\PT$, we run additional simulations to tune the coefficients $C_{0\PN}$ and $C_{1\PN}$.
We construct polynomial fits in the ordinary and symmetric mass ratio to allow the hybrid waveform to be evaluated for mass ratios as small as $1/8$.
We describe the calibration steps for each process in more detail below.
The result of this process are waveforms that have an $O(10^{-3})$ error measure computed from the expression in Eq.~\eqref{eq:cal_e}.

\subsection{\label{sec:ModifiedPT} Modified-potential calibration}

The first stage in the calibration process is to tune the parameters $V_\PT$ and $\kappa_\PT$ so that the least-damped QNM mode of the potential matches that in the $l=2$, $m=\pm 2$ modes of the ringdown waves of the NR waveform.
We performed this optimization of the potential for each mass ratio by running simulations with different values of $\kappa_\PT$ and $V_\PT$ with the parameters $C_{0\PN}$ and $C_{1\PN}$ set equal to one and zero, respectively.
Specifically, we first generate waveforms with the hybrid method for a 2D grid of values of $\kappa_{\mathrm{PT}}$ and $V_{\mathrm{PT}}$ for each mass ratio that is used in the calibration of the hybrid method.
Next, we determine the corresponding QNM frequencies (both real and imaginary parts $\omega = \omega_\mathrm{R} + i \omega_\mathrm{I}$) from the waveforms generated from the hybrid method for the different $\kappa_{\mathrm{PT}}$ and $V_{\mathrm{PT}}$ values used at each mass ratio. 
We do this by fitting a single damped exponential function $\mathcal A e^{i\omega t}$ to the data on the order of $10M$ after the peak of the hybrid waveform (with $\mathcal A$ and $\omega$ being the coefficients in the fitting function). 

Using the $\omega_\mathrm{R}$ and $\omega_\mathrm{I}$ extracted from each simulation of the hybrid-method waveform for different $\kappa_{\mathrm{PT}}$ and $V_{\mathrm{PT}}$ values, we construct interpolation functions $\omega_\mathrm{I}(\kappa_{\text{PT}}, V_{\text{PT}})$ and $\omega_\mathrm{R}(\kappa_{\text{PT}}, V_{\text{PT}})$.  
By setting these interpolation functions equal to the least-damped QNM consistent with the final mass and spin of the remnant black hole in the NR simulation (namely, $\omega^{\mathrm{NR}}_\mathrm{I}$ and $\omega^{\mathrm{NR}}_\mathrm{R}$), we determine the optimal values of $\kappa_{\mathrm{PT}}$ and $V_{\mathrm{PT}}$.
Finally, we perform a polynomial fit to the optimized $\kappa_{\mathrm{PT}}$ and $V_{\mathrm{PT}}$ as a function of the symmetric mass ratio $\nu$.
The expressions for these fitting functions are given by
\begin{subequations} \label{eq:kappa_vpt_fit}
\begin{align} 
M^2V_\mathrm{PT}(\nu) = {} & 9.50928\,\nu^3 - 2.74236\,\nu^2 + 0.891781\,\nu  \nonumber \\
& + 0.121403 \,, \\
M\kappa_\mathrm{PT}(\nu) = {} & -16.2354\,\nu^3 + 7.8138\,\nu^2- 1.5761\,\nu  \nonumber \\
&  + 0.2709\, .
\end{align}
\end{subequations}
Plots of the fitting functions and the data used to produce these functions are shown in Appendix~\ref{app:Fitting}. 

\begin{figure}[htb]
    \centering
    \includegraphics[width=\columnwidth]{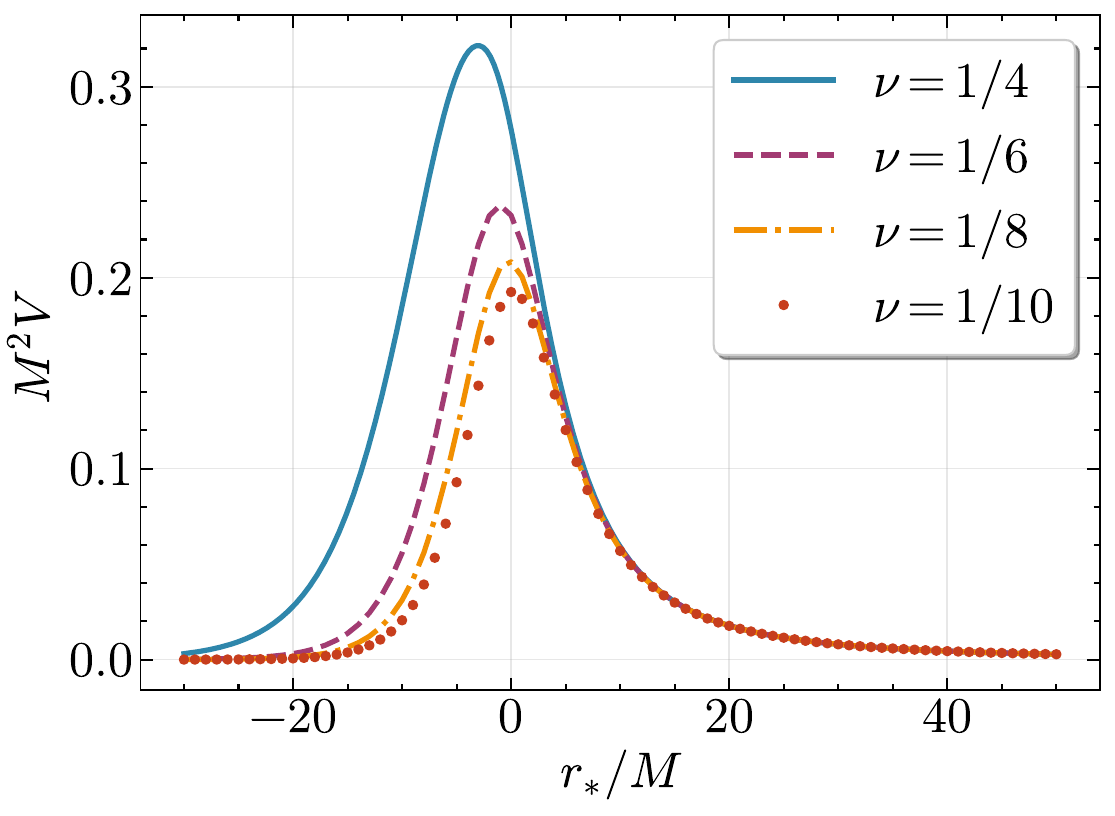}
    \caption{\textbf{Calibrated modified Poschl-Teller potential for different mass ratios}. Specifically, the solid blue curve represents the mPT potential for $\nu = 1/4$ (equal mass), the dashed purple curve is for $\nu = 1/6$, the dash-dotted orange is for $\nu = 1/8$ and the red dotted curve represents $\nu = 1/10$.
    As described in more detail the text of Sec.~\ref{sec:ModifiedPT}, the potential decreases in amplitude and width as the symmetric mass ratio decreases.}
    \label{fig:pot_vs_nu}
\end{figure}

We conclude this section by showing how the mPT potential (for the optimized $\kappa_{\mathrm{PT}}$ and $V_{\mathrm{PT}}$ values) varies as a function of the mass ratio. 
This is illustrated in Fig.~\ref{fig:pot_vs_nu}, which plots four of the five mass ratios used in the calibration of the hybrid method.
As the final spin of the black hole decreases for decreasing symmetric mass ratio, the height and width of the mPT potential also decrease.
These trends in the potential as a function of mass ratio correspond with the fact that the oscillation frequency and damping timescale of the least-damped quasinormal mode both decrease with decreasing spin parameter.

\subsection{\label{sec:Trajectory} Boundary-data calibration}

After determining the optimal values of $\kappa_{\mathrm{PT}}$ and $V_{\mathrm{PT}}$, we can now fix these values and optimize the values of $C_{0\PN}$ and $C_{1\PN}$, which we added to the boundary data---see Eq.~\eqref{eq:zerilli_shell}. 
Our prescription for how we optimize the $C_{0\PN}$ and $C_{1\PN}$ coefficients differs from that of the potential parameters $\kappa_{\mathrm{PT}}$ and $V_{\mathrm{PT}}$; we will now describe our procedure.

Because coefficient $C_{0\PN}$ corresponds to an overall rescaling of the boundary data, and because the Zerilli function is a linear partial differential equation, we find that changing the amplitude of $C_{0\PN}$ leads to a proportional rescaling of the gravitational waveform.
The term in the boundary data that $C_{1\PN}$ multiplies, however, scales as $M/R$, so its effect on the boundary data (and the waveform generated by this boundary data) becomes more significant when $R$ is smaller (or at times corresponding to the late inspiral and the merger).
These properties influence how we calibrate these parameters, as we now describe.

In the discussion below, we use the convention that the peak of the $l=2$, $m=2$ spherical harmonic mode takes place at time $t=0$ (as in the NR surrogate models).
We first choose a time during the earlier stages of the inspiral (specifically $t/M = -1500$), where we scale the boundary data by a value of $C_{0\PN}$ so that the hybrid waveform agrees with the NR at that time.
Since the effects of $C_{1\PN}$ are strongest on the waveform during the late stages of the inspiral, we calibrate the coefficient by running a sequence of shorter duration simulations (starting at $t/M = -500$) for different values of $C_{1\PN}$ for each of the five calibration mass ratios. 
To assess the best fit value of the $C_{1\PN}$ parameter, we minimize the error in Eq.~\eqref{fig:error} between the hybrid and NR waveforms as a function of $C_{1\PN}$.
While the effect of $C_{1\PN}$ is small during the earlier inspiral, we do find that it can make enough of a change that we find it necessary to rescale the waveform by readjusting the value of $C_{0\PN}$ after the optimization of $C_{1\PN}$.
We have checked that optimizing the $C_{0\PN}$ coefficient at the arbitrary time of $t/M=-1500M$ still produces good agreement of the hybrid and NR waveforms at a thousand $M$ earlier in the inspiral. 

With the coefficients $C_{0\PN}$ and $C_{1\PN}$ that were optimized to match NR waveforms at five mass ratios, we next construct polynomial fits to these coefficients, so that we can have an interpolating function that we can evaluate for any symmetric mass ratio between $\nu = 1/4$ and $1/10$ (as we did with $\kappa_{\mathrm{PT}}$ and $V_{\mathrm{PT}}$).
Given the dependence of the data on $\nu$ and $q$, we found that a polynomial fit in $q$ better matched the data for $C_{0\PN}$, whereas the polynomial fit in $\nu$ was a better match for $C_{1\PN}$.
The respective fitting functions are given by
\begin{subequations} \label{eq:CPNs}
\begin{align} \label{eq:c0pn_eqn}
C_{0\PN}(q) = &  -0.616335\,q^3 + 1.522547\,q^2 - 1.549295\,q \nonumber \\
& + 0.867041\,, \\
\label{eq:C1PNfit}
C_{1\PN}(\nu) =  & -34.1631\,\nu^3 + 36.6120\,\nu^2 - 10.1504\,\nu \nonumber \\
& + 0.6832 \nu  + 1.06181\, .
\end{align}
\end{subequations}
The fitting functions are plotted in Appendix~\ref{app:Fitting} with the optimized values of the parameters at the five mass ratios.
We also note that the fit for $C_{0\PN}$ can also be viewed as a fit in $\nu$ in a form that is not polynomial by writing $q$ in terms of $\nu$ as
\begin{align} \label{qvsnu_eqn}
q = \frac 1{2\nu} \left(1 - 2\nu - \sqrt{1 - 4\nu} \right) .
\end{align}

\section{\label{sec:GW} Results and comparison with numerical relativity}

In this section, we compare the waveform generated using the hybrid method (with the modifications described in Sec.~\ref{sec:modifications} after being calibrated as described in Sec.~\ref{sec:calibration}) to corresponding waveforms generated by NR surrogate models.

We perform simulations with the hybrid method using the characteristic evolution algorithm in~\cite{Gundlach:1993tp} (see also~\cite{Nichols:2010qi,Nichols:2011ih} for the treatment of the boundary data).
We use a time step of $\delta u/M = \delta v/M = 1/2$, solve the partial differential equation (PDE) on a computational domain of size roughly $10^4 M$ in the null coordinates $u$ and $v$.
This allows us to obtain waveforms that are thousands of $M$ long after junk radiation leaves the computational grid (the reason for the junk radiation in the hybrid method is discussed in~\cite{Nichols:2010qi,Nichols:2011ih}).
We extract the waves at a tortoise coordinate of $200M$, because we found empirically that extracting at larger radii does not make significant changes in the hybrid-method waveform.

For the potential, we use the fitting functions for $V_\mathrm{PT}$ and $\kappa_\mathrm{PT}$ in Eq.~\eqref{eq:kappa_vpt_fit} to determine the relevant coefficients in our mPT potential.
We also use the fitting functions in Eq.~\eqref{eq:CPNs} for the coefficients $C_{0\PN}$ and $C_{1\PN}$ that enter into the boundary data.
Using the calibrated data, solving the characteristic evolution equations (a 2D system of PDEs) takes on the order of a few minutes on one core of a laptop computer.

The solution of the PDE system using the hybrid method produces the gauge-invariant Zerilli function $\Psi$ throughout the computational domain.
Asymptotically, the gauge-invariant Zerilli function (expanded in scalar spherical harmonics) can be related to $r$ times the spin-weighted spherical-harmonic moments of the gravitational-wave strain $h = h_+-ih_{\times}$ (see, e.g.,~\cite{Martel:2005ir}).
For the $l=2$ modes, the constant of proportionality is $\sqrt 6$ and the relation is given by 
\begin{equation}
    \lim_{r\rightarrow\infty} \Psi_{2,2} = \frac{1}{\sqrt 6} r h_{2,2}
\end{equation}
We use this relationship to compare with the NR surrogate waveforms.

\begin{figure*}
    \centering
    \includegraphics[width=\textwidth]{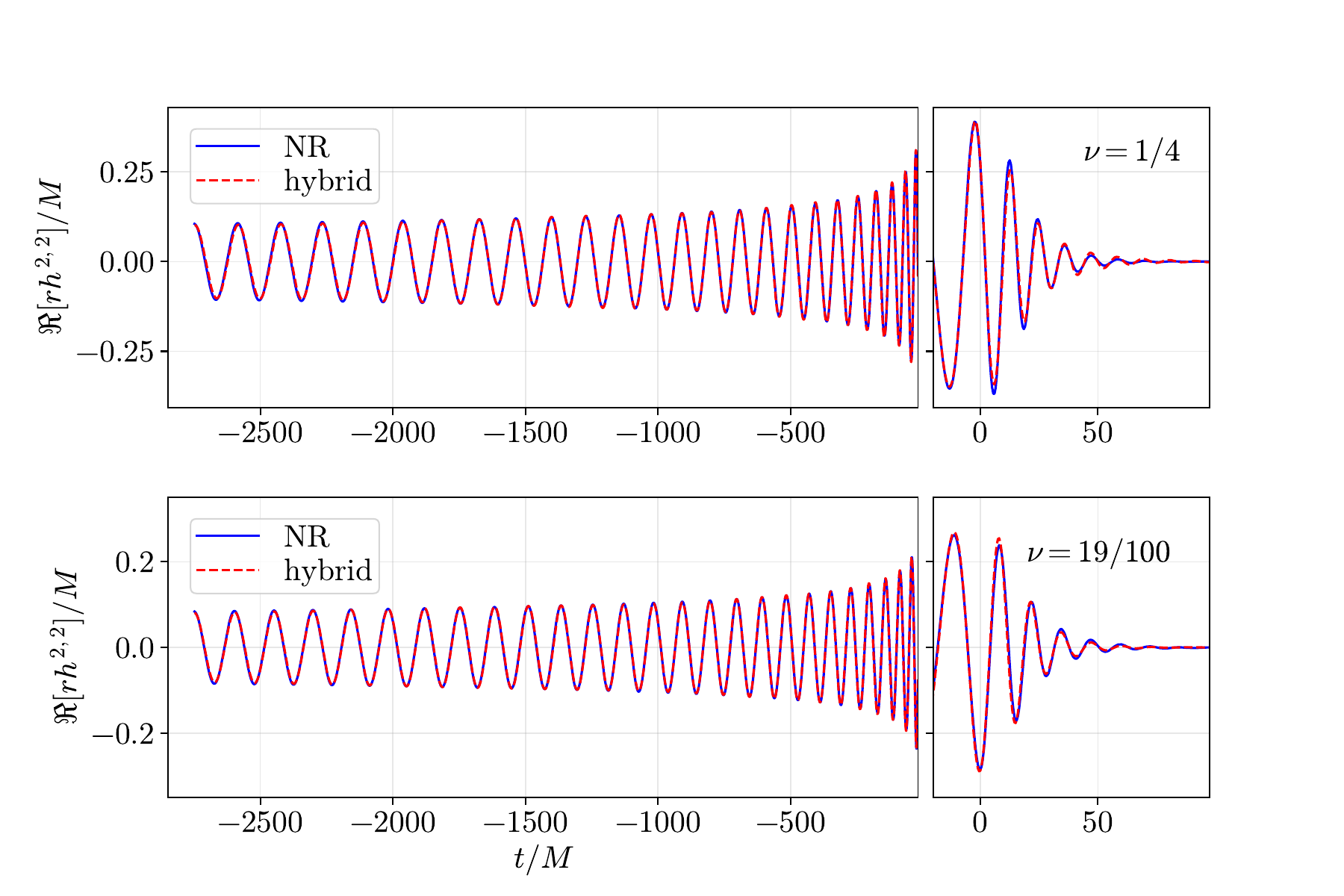}
    \caption{\textbf{Comparison of the hybrid-method and NR waveforms}:
    In both panels, the red dashed curves are the waveform produced by the hybrid method, and the solid blue curves are the NR waveforms. 
    The top and bottom panels correspond to waveforms for mass ratios $\nu=1/4$ and $\nu=19/100$, respectively. 
    The right side of each panel zooms in on the merger and ringdown stages of the waveform, which change on a shorter timescale than that of the inspiral shown on the left.}
    \label{fig:hyb_nr2}
\end{figure*}

Figure~\ref{fig:hyb_nr2} shows that the full IMR waveforms computed using the hybrid method as red-dashed curves and those from the NR surrogate as solid blue curves.
The top panel of the figure is for the symmetric mass ratio $\nu=1/4$ and the lower panel shows $\nu=19/100$. 
These two panels show (visually) that the phenomenological modifications that we made to the hybrid method in this paper lead to significantly better agreement with NR waveforms than the results in~\cite{Nichols:2010qi,Nichols:2011ih} did, which did not make use of any calibration.

\begin{figure}[htb]
    \centering
    \includegraphics[width=\columnwidth]{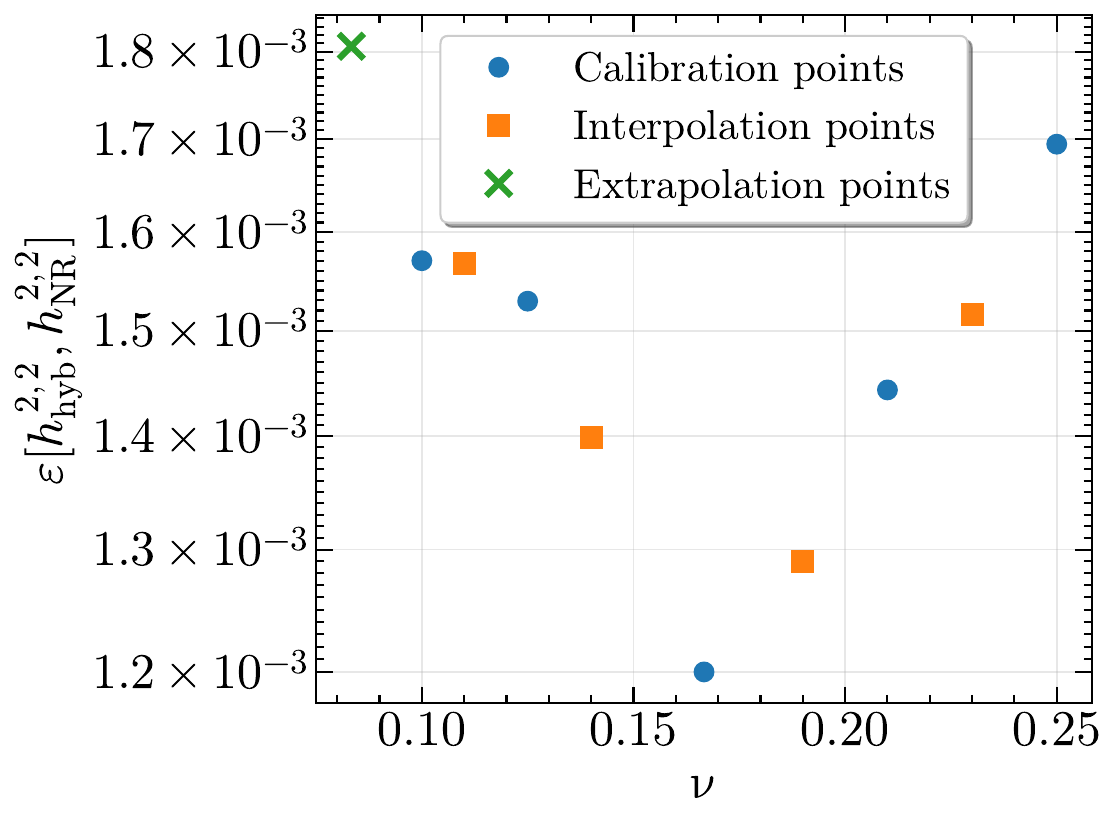}
    \caption{\textbf{Waveform error versus mass ratio}:
    The Normalized $L^2$ errors, as defined in Eq.~\eqref{eq:cal_e}, between the hybrid and NR waveforms for several mass ratios between $\nu=1/4$ and $\nu=1/12$. 
    The symmetric mass ratios $\nu = 1/4, 21/100, 1/6, 1/8, 1/10$ (shown as filled blue circles) were used to calibrate the fitting functions used in the hybrid method. 
    The symmetric mass ratios $\nu = 23/100, 19/100, 14/100$ and $11/100$ (the orange squares) were used to check how well the hybrid method performs at mass ratios within the domain of the interpolation function, whereas the symmetric mass ratio $\nu = 1/12$ (the green ``$\times$'' marker) was generated using the hybrid method by extrapolating the fitting functions.
    The error measure is of order $10^{-3}$ for all the mass ratios shown.}
    
    \label{fig:error}
\end{figure}

To make a more quantitative assessment of the agreement of the hybrid and NR waveforms, we also show in Fig~\ref{fig:error} the values of the error measure in Eq.~\eqref{eq:cal_e} for specific mass ratios from $\nu = 1/4$ to $\nu = 1/12$. 
The values of the error at the mass ratios used to calibrate the fitting functions are the blue circles.
The orange squares between these points are the error at mass ratios that fall between the calibration points, whereas the point at a mass ratio at $\nu = 1/12$ is computed by running the hybrid by extrapolating our fitting function outside of the values of mass ratio that were used to calibrate our model.
In all cases (calibration, interpolation, and extrapolation) the errors are on the order of $10^{-3}$.

\section{\label{sec:conclusions} Conclusions}

In this paper, we adapted the hybrid method of generating approximate IMR waveforms~\cite{Nichols:2010qi,Nichols:2011ih} to better agree with the IMR waveforms produced by NR simulations.
To make these improvements, while still capturing the qualitative features of the hybrid method---namely, that the features of a full IMR waveform can be generated through the interaction of post-Newtonian-inspired boundary data in a black-hole perturbation problem---, we found that it was necessary to introduce several phenomenological modifications to the hybrid method.

The first significant change was that we evolved the boundary region by assuming that it was proportional to the orbital separation of the binary within the EOB framework.
This change was necessary to improve the agreement of the NR and hybrid-method waveforms during the inspiral.
The EOB dynamics are typically truncated during the plunge, but we require that our boundary region continue for all retarded times in the hybrid-method simulation.
Thus, we generated a phenomenological extension of the EOB trajectory which smoothly attached to the trajectory prior to the plunge, so that the merger and ringdown stages of the hybrid waveform would not be affected.
Our prescription for constructing the trajectory and the corresponding boundary region did not require any further calibration against NR simulations.

A second significant change to the hybrid method arose from modifying the effective potential of the black-hole perturbation problem, so as to obtain better agreement between the ringdown waves in NR and in the hybrid method.
Similarly to how the EOB method models an effective Hamiltonian dynamics of the two-body problem (supplemented by a radiation-reaction generalized force) as a one-body problem on an effective spacetime, here we modified the effective potential of a Regge-Wheeler-Zerilli black-hole perturbation theory problem to produce a least-damped QNM that better matches the equivalent QNM generated in NR simulations of binary black holes.
We could calibrate the two free parameters in our ansatz for the potential's form to produce a waveform that agrees better during the ringdown stage.

A more minor change to the hybrid method arose in the boundary data.
Although we continued to use the leading, Newtonian-order expression for the Zerilli function on the boundary, we introduced two calibration parameters that modified the overall amplitude of the boundary data and introduced a PN-inspired radial variation of the boundary data that helped improve the match between the hybrid waveform and the NR waveforms during the late inspiral and merger.
Tuning these parameters was the final calibration stage of the hybrid method.

After calibrating the hybrid method against five waveforms from nonspinning BBH systems with different mass ratios, we obtained fitting functions that allowed the hybrid method to generate waveforms for nonspinning BBH systems with mass ratios between one and one-eighth.
The accuracy of the hybrid waveform model was on the order of $10^{-3}$ using the waveform error defined in Eq.~\eqref{eq:cal_e}.
Our results show that the picture of the full waveform as being generated from boundary data in an effective linear black-hole perturbation spacetime does capture the full IMR waveform from NR simulations to good approximation.  

There are a few possible avenues to generalize the work here.
First, it would be of interest to use the hybrid method to model the waveform from binaries with spins (anti-)aligned with the orbital angular momentum, or even for generic spin orientations.
Whether this would require significant additional modifications to the hybrid (or just more extensive calibration of the method developed in this paper) would be of interest to determine.
A similar statement also holds for considering more extreme mass ratios.

There have also been an increasing number of NR simulations performed in beyond-GR gravitational theories (see e.g.~\cite{Okounkova:2017yby,Okounkova:2019zjf,Okounkova:2020rqw,Silva:2020omi,East:2020hgw,East:2021bqk,Corman:2022xqg,Corman:2022xqg,Corman:2024vlk,Kuan:2023hrh,Kuan:2023trn}).
These theories frequently have extra scalar or tensor degrees of freedom that can couple nonminimally to gravity and produce additional polarizations of gravitational waves.
Extending the hybrid method to include these additional fields and polarizations of gravitational waves would also be of interest.

\acknowledgments
N.R.\ and K.Y.\ acknowledge support from the NSF grant PHY-2309066 and the Owens Family Foundation.
D.A.N.\ acknowledges support from the NSF grants PHY-2011784 and PHY-2309021. 
D.A.N.\ is also supported by the NSF-CAREER Award PHY-2439893 and K.Y.\ is also supported by the NSF-CAREER Award PHYS-2339969.

% --------------------------------------------------------------------------------------

\appendix

\section{\label{app:Fitting} Fitting functions for the calibration parameters}

\begin{figure*}[htb]
    \centering
    \includegraphics[width=\textwidth]{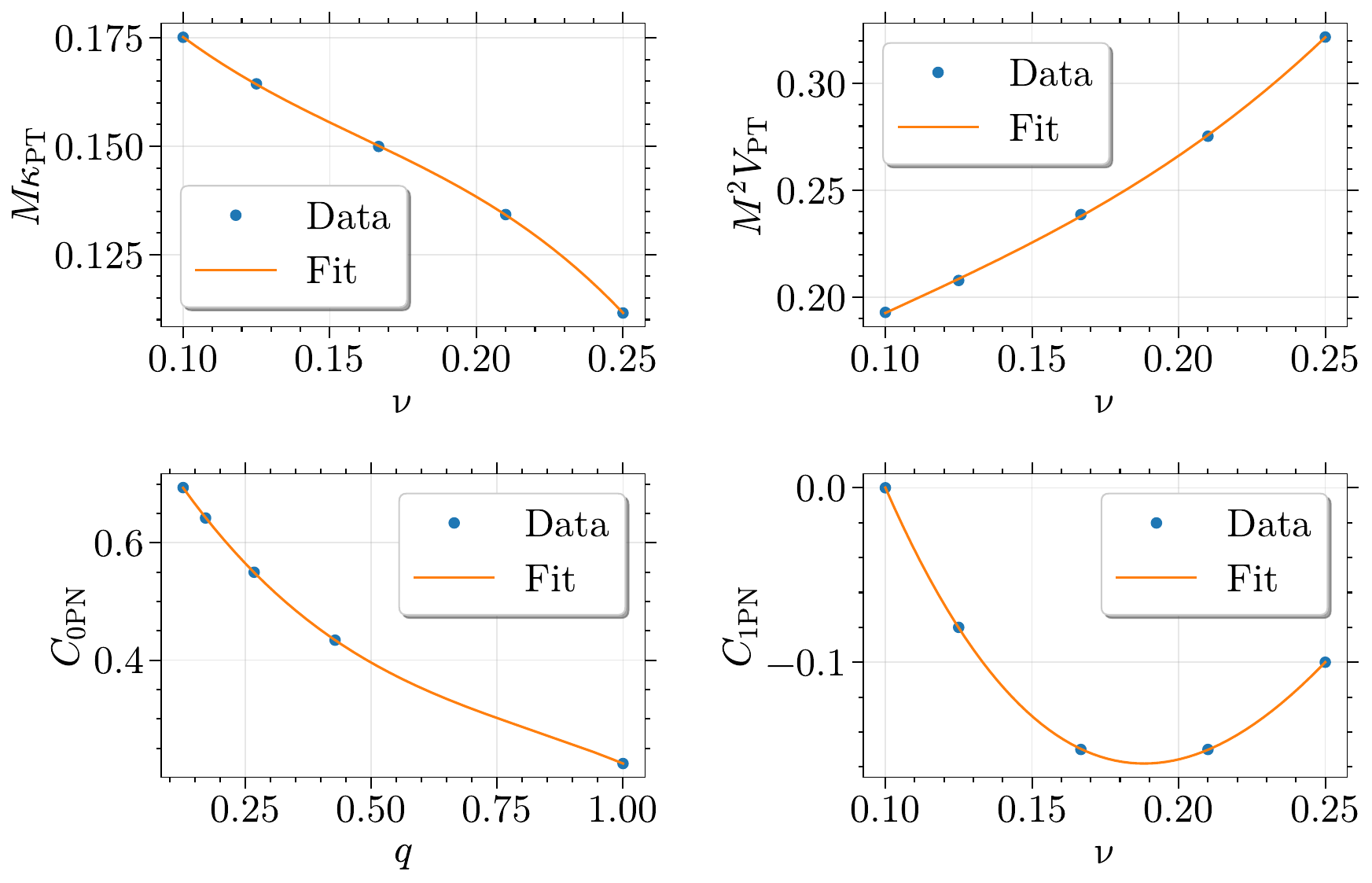}
    \caption{Comparison of the data and fit for $M \kappa_\mathrm{PT}$ versus $\nu$ (top left), $M^2 V_\mathrm{PT}$ versus $\nu$ (top right),  $C_\mathrm{0PN}$ versus $q$ (bottom left), and $C_\mathrm{1PN}$ versus $\nu$ (bottom right). 
    The corresponding fits are given in Eqs.~\eqref{eq:kappa_vpt_fit} for the top left and right, and Eq.~\eqref{eq:CPNs}, for the bottom left and right panels, respectively.}
    \label{fig:allfit}
\end{figure*}

In this appendix, we present a few supplementary results related to the calibration results of the modified PT potential and the coefficients associated with the boundary data ($C_{0\PN}$ and $C_{1\PN}$).

The main results are shown in Fig.~\ref{fig:allfit}.
They show the four parameters that we optimize in the hybrid method ($\kappa_\PT$, $V_\PN$, $C_{1\PN}$ and $C_{0\PN}$) in the four panels clockwise from the top left.
The blue dots labeled in the legend as ``data'' are the optimized values at the five mass ratios used to calibrate the hybrid method.
The orange curves are the cubic fits which were given in Eqs.~\eqref{eq:kappa_vpt_fit} and~\eqref{eq:CPNs}.
As discussed in more detail in Sec.~\ref{sec:Trajectory}, the fit for $C_{0\PN}$ is better fit by a polynomial in $q$, whereas the other coefficients are better fit by a polynomial in $\nu$.

\bibliography{References}

\end{document}